\documentclass[aps,prl,superscriptaddress,amsmath,amssymb,floatfix,twocolumn,longbibliography]{revtex4-1}
\usepackage{graphicx}
\usepackage{subfigure}
\usepackage{color}
\usepackage{epstopdf}
\usepackage{amsmath}
\usepackage{bm}
\usepackage{ulem}
\usepackage{comment}

\begin{document}
\title{Orbital-selective electron correlations in high-$T_{\rm c}$ bilayer nickelates:
from a global phase diagram to implications for spectroscopy}

\author{Zhiguang Liao}
\thanks{These authors contributed equally to this study.}
\affiliation{Department of Physics and Beijing Key Laboratory of Opto-electronic Functional Materials \& Micro-nano Devices,
Renmin University of China, Beijing 100872, China}

\author{Yiming Wang}
\thanks{These authors contributed equally to this study.}
\affiliation{Department of Physics \& Astronomy,
Extreme Quantum Materials Alliance, Smalley Curl Institute,
Rice University, Houston, Texas 77005,USA}

\author{Lei Chen}
\affiliation{Department of Physics \& Astronomy,
Extreme Quantum Materials Alliance, Smalley Curl Institute,
Rice University, Houston, Texas 77005,USA}
\affiliation{Department of Physics and Astronomy, Stony Brook University,
Stony Brook, NY 11794, USA}

\author{Guijing Duan}
\affiliation{Department of Physics and Beijing Key Laboratory of Opto-electronic Functional Materials \& Micro-nano Devices,
Renmin University of China, Beijing 100872, China}

\author{Rong Yu}
\email{rong.yu@ruc.edu.cn}
\affiliation{Department of Physics and Beijing Key Laboratory of Opto-electronic Functional Materials \&
Micro-nano Devices, Renmin University of China, Beijing 100872, China}
\affiliation{Key Laboratory of Quantum State Construction and Manipulation (Ministry of Education),
Renmin University of China, Beijing, 100872, China}

\author{Qimiao Si}
\email{qmsi@rice.edu}
\affiliation{Department of Physics \& Astronomy,
Extreme Quantum Materials Alliance, Smalley Curl Institute,
Rice University, Houston, Texas 77005,USA}

\begin{abstract}
Motivated by the high temperature superconductivity observed in the bilayer nickelate
La$_3$Ni$_2$O$_7$ and the spectroscopic evidences of strong electron correlations in this compound,
we address the role of its multiorbital electron correlations
by proposing a global phase diagram of a bilayer two-orbital Hubbard model.
We find a Mott transition developing at half filling, and identify strong orbital selectivity when the system is
at the physical electron count. The orbital selectivity is manifested in the formation of interlayer spin singlets between
electrons in the $z^2$ orbitals. These features lead to a strong renormalization of the electronic band structure
while sustaining a sizable splitting between the bonding and antibonding $z^2$ bands.
The proposed orbital-selective correlations naturally explain a series of features as observed
in the angular resolved photoemission spectroscopy (ARPES) and optical conductivity measurements in La$_3$Ni$_2$O$_7$.
Our results provide a basis to understand both the normal state and the high temperature superconductivity of
multilayer nickelates and thereby elucidate correlated superconductivity in general.
\end{abstract}

\maketitle


\textit{Introduction.~}The discovery of superconductivity about 80 K in the
bilayer nickelate La$_3$Ni$_2$O$_7$ under high pressure~\cite{Sun_Nature_2023} is of extensive current interest.
Notwithstanding concerns on the structural
imperfection~\cite{arXiv:2312.15727Chen, 1313, arXiv:2311.12361Sun},
evidence for the bulk
nature of the superconductivity
is growing~\cite{Yuan_NP_2024, arXiv:2407.05681Cheng},
and
superconductivity has
very recently been seen in thin films
at ambient pressure~\cite{Hwang_Nature_2024}.
While their phenomenology, with superconductivity developing upon the suppression of
an electronic order,
bears some similarities with
other families of correlated superconductors, the microscopic electronic physics of the bilayer nickelates
has some clear distinction.
Unlike the infinite-layer nickelate (Sr,Nd)NiO$_2$ thin films~\cite{LiHwang_Nature_2019},
which
parallels the high-$T_{\rm c}$ cuprates~\cite{Lee_RMP_2006}
in that it shares the latter's 3d$^9$ electron configuration from
the Ni$^{+}$ ion,
La$_3$Ni$_2$O$_7$ has a valence count of Ni$^{+2.5}$, corresponding to
the electron configuration
3d$^{7.5}$.
The implied multiplicity of the active $3d$ electrons
is reminiscent of
the Fe-based
superconductors~\cite{YuAbrahamsSi_NatRevMater_2016, Bohmer_NP_2022, SiHussey_PT_2023}.
{\it Ab initio} calculations further suggest that the $z^2$ and $x^2-y^2$
3d orbitals are important to the low-energy electronic structure~\cite{Yao_PRL_2023}.
Accordingly, there are $N=3$ electrons per bilayer unit cell that are active in the manifold of Ni $e_g$ orbitals.
All these point towards the importance of orbital-selective correlations
in La$_3$Ni$_2$O$_7$, as we have emphasized in Ref.~\cite{Liao_PRB_2023}.
Other theoretical approaches
to the electron correlations have also been
taken~\cite{ShilenkoLeonov_PRB_2023,
LechermannEremin_PRB_2023, ZhangDagotto_PRB_2023, CaoYang_PRB_2024,
OuyangLu_PRB_2024, RyeeWehling_PRL_2024, WangHu_arXiv_2024}.
More generally,
 the origin of
the high-temperature superconductivity in La$_3$Ni$_2$O$_7$
has been the focus of enormous theoretical
efforts~\cite{QuSu_PRL_2024, HeierSavrasov_PRB_2024,
ZhanHu_arXiv_2024, TianLu_PRB_2024, PanWu_arXiv_2023, ChangLi_arXiv_2023,
WangYang_arXiv_2024, QinYang_PRB_2023, LuoYao_npjQM_2024, JiangZhang_CPL_2024,
HuangZhou_PRB_2023, YangZhang_PRB_2023, LuWu_PRL_2024, LuWu_PRB_2024,
XueWang_CPL_2024, ChenLi_PRB_2024, KanekoKuroki_PRB_2024, KakoiKuroki_PRB_2024,
MaWu_arXiv_2024, Yang_arXiv_2024, SakakibaraKuroki_PRL_2024, YangWang_PRB_2023,
JiangKu_PRL_2024, LiuChen_arXiv_2023, QuSu_arXiv_2023, YangZhang_arXiv_2024,
YangZhang_arXiv_2023, ZhangWeng_PRL_2024, LuYou_arXiv_2023, FanXiang_PRB_2024,
ZhengWu_arXiv_2023, SchlomerBohrdt_arXiv_2023, BotzelEremin_arXiv_2024_1,
BotzelEremin_arXiv_2024_2, OhZhang_arXiv_2024, OhZhang_PRB_2023, Dagotto_PRB_2023, Dagotto_2024}.

The effects of electron correlations in bilayer nickelates have been studied in experiments at ambient pressure.
Four salient features have emerged.
First, a recent measurement of optical conductivity~\cite{LiuWen_NC_2024} reported a substantially suppressed
Drude weight compared to the band theory value.
Second, an ARPES experiment~\cite{YangZhou_NC_2024} found strong orbital-dependent band renormalizations,
suggesting orbital-selective electron correlations in the system. While both points
indicate substantial band renormalization effect, the spectroscopy measurements provide two further
constraints on the theoretical description. As the third point, the interband peak
of the optical spectra,
located at about 1 eV, seems to exhibit
relatively weak renormalization effect compared to the density functional theory (DFT) results~\cite{LiuWen_NC_2024, GeislerHirschfeld_npjQM_2024}.
The fourth point concerns whether or not the z$^2$ bonding band (the $\gamma$ band) crosses the Fermi level ($E_F$).
This influences the Fermi surface topology and is possibly important to the superconductivity at high pressure.
The aforementioned ARPES
experiment revealed that this band
is located about 50 meV below the Fermi level and its energy position is almost temperature independent,
implying
that this
gap to $E_F$ is unrelated to the density wave ordering.
Qualitatively, a strong band renormalization effect could enable such a
gap; however, it would seem that such an effect is incompatible with
the
weak renormalization of the
observed interband optical conductivity peak.
Taken together, these four points represent a challenge to a comprehensive
theoretical description.

\begin{figure}
\includegraphics[width=0.45\textwidth]{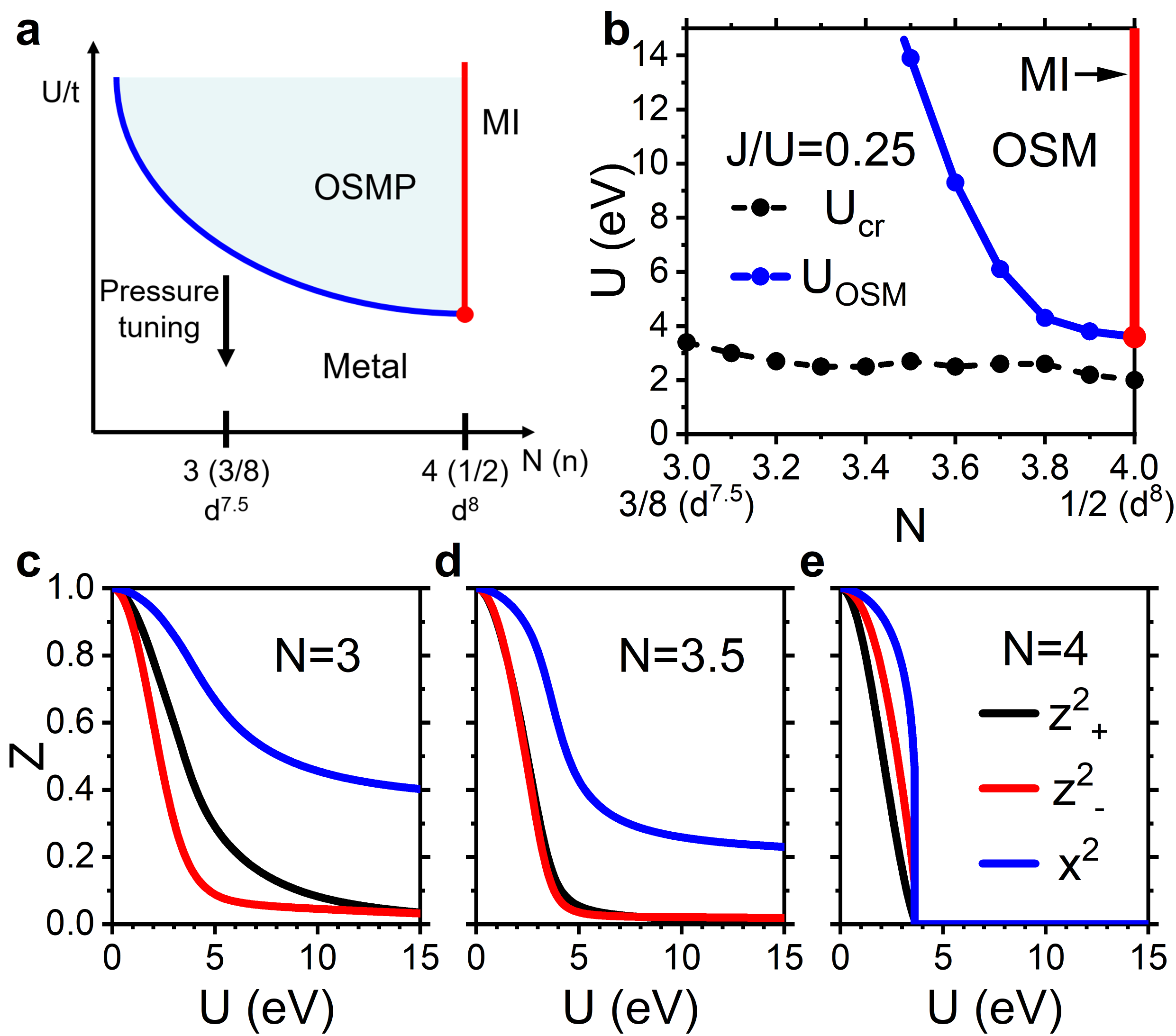}
\caption{{\bf a}
    The proposed ground-state phase diagram with $U$ and electron occupation
	number at nonzero Hund's coupling $J_{\text{H}}$ of the bilayer two-orbital
	Hubbard model for La$_3$Ni$_2$O$_7$,
	anchored by the calculations shown in {\bf b}-{\bf e}.
	The red line
	denotes a Mott insulator (MI)
	 where both orbitals are
	localized. The shaded regime away from half-filling stands for an
	orbital-selective Mott phase (OSMP) where electrons in the $z^2$ orbital
	are localized while those in the $x^2-y^2$ orbital remain itinerant. The
	system
	exhibits strong orbital-selective behavior even outside
	(though still in proximity of)
	the OSMP. Accordingly, the black arrow marks the parameter regime for La$_3$Ni$_2$O$_7$.
	The electron filling factors and configurations are marked under the 
	horizontal axis. {\bf b} Calculated phase diagram 
	of the bilayer two-orbital model at $J_{\rm{H}}/U=0.25$. The 
	system undergoes a Mott transition to a MI by increasing 
	$U$ at $N=4$, corresponding to half filling. The blue line denotes the 
	phase boundary of the OSMP away from  
	half filling. The dashed black 
	line indicates a crossover to a strongly orbital-selective metallic phase 
	by increasing $U$.
	{\bf c}-{\bf e}
	Evolution of the orbital-resolved quasiparticle spectral weight $Z$ with $U$ in the bilayer two-orbital Hubbard model for $J_{\text{H}}/U=0.25$ and at electron numbers $N=3$, $N=3.5$, and $N=4$, respectively,
    showing the strong orbital-selective bahavior and,
    for $N=4$, the Mott transition.
}
\label{Fig:1}
\end{figure}

In this paper, we provide a coherent understanding of all these points in a 
bilayer two-orbital
Hubbard model for La$_3$Ni$_2$O$_7$ in terms of orbital-selective electron correlations,
which we characterize through a proposed global phase diagram
of the Coulomb interaction strength ($U$) and electron density ($N$) as given in Fig.~\ref{Fig:1}{\bf a}.
We show that a Mott transition develops
with increasing
$U$ when the electron density
is taken to be
at half filling
($N=4$); the resulting
Mott insulator anchors
the strong orbital-selective behavior
of the system when $N$ moves
away from half filling up to the physical electron count $N=3$.
In addition,  the strong orbital selectivity causes an effective
interlayer antiferromagnetic (AFM) exchange interaction between electrons in
the $z^2$ orbital. The AFM interaction further bounds the electrons between the
top and bottom layers to an interlayer spin singlet and this causes an additional splitting
between the bonding and antibonding $z^2$ bands.
By calculating the band structure and the optical conductivity in the model,
we show that our theory provides the understanding
of the strong orbital-selective band renormalization and the sinking of the
$\gamma$ band,
along with
the substantially reduced Drude weight
and yet the weakly shifted interband peak in the optical conductivity.


\textit{Model and method.~}We consider a bilayer
Hubbard model
of two orbitals,
corresponding to the 3d $z^2$ and $x^2-y^2$ orbitals.
The Hamiltonian reads as
$H = H_{\rm{TB}} + H_{\rm{int}}$. The tight-binding Hamiltonian is as follows:
\begin{eqnarray}
\label{Eq:Ham_0}
H_{\rm{TB}} && =\frac{1}{2}\sum_{i\delta ll^\prime\alpha\beta\sigma}
t^{\alpha\beta}_{\delta ll^\prime}
d^\dagger_{il\alpha\sigma} d_{i+\delta l^\prime\beta\sigma} 
 + \sum_{il\alpha\sigma} (\epsilon_\alpha-\mu)
d^\dagger_{il\alpha\sigma} d_{il\alpha\sigma}
\, .
\nonumber\\
\end{eqnarray}
Here, $d^\dagger_{il\alpha\sigma}$ creates an electron in orbital $\alpha$
with spin $\sigma$ at site $i$ of the square lattice in layer $l$
(with $l=1,2$ denoting the top and bottom layers, respectively),
$\epsilon_\alpha$
refers to the energy level associated with the crystal field
splittings,
and $\mu$ is the chemical potential.
The tight-binding parameters can be found in the Supplemental Materials
(SM)~\cite{SM}.
In addition, the interacting Hamiltonian contains the dominant on-site interactions:
\begin{eqnarray}\label{Eq:Ham_int}
H_{\rm{int}} &=&
\frac{U}{2}
\sum_{i,l,\alpha,\sigma}n_{il\alpha\sigma}n_{il\alpha\bar{\sigma}} \nonumber\\
&+&
\sum_{i,l,\alpha<\beta,\sigma} \left
\{ U^\prime n_{il\alpha\sigma}
n_{il\beta\bar{\sigma}} \right.
+ (U^\prime-J_{\rm{H}}) n_{il\alpha\sigma} n_{il\beta\sigma}
\\
&-&\left. J_{\rm{H}}(d^\dagger_{il\alpha\sigma}d_{il\alpha\bar{\sigma}}
d^\dagger_{il\beta\bar{\sigma}}d_{il\beta\sigma}
+d^\dagger_{il\alpha\sigma}d^\dagger_{il\alpha\bar{\sigma}}
d_{il\beta\sigma}d_{il\beta\bar{\sigma}}) \right\} \, , \nonumber
\end{eqnarray}
where $n_{il\alpha\sigma}=d^\dagger_{il\alpha\sigma} d_{il\alpha\sigma}$.
Here,
$U$, $U^\prime$, and $J_{\rm{H}}$, respectively denote the intra-
and inter- orbital repulsions and the Hund's rule coupling,
with
$U^\prime=U-2J_{\rm{H}}$
taken~\cite{Castellani_PRB_1978}.

We study the model by the
U(1)
slave-spin method, which incorporates electron correlation effects such as 
the renormalization of electronic structures and dynamical spectral weight 
transfer, 
as detailed in Refs.~\cite{Yu_PRB_2012, Yu_PRB_2017, 
		YuSi_PRB_2011}.
To capture the effects of the strong interlayer hopping in the
$z^2$ orbital, we perform the calculation in a two-site unit cell and rotate
the atomic
basis
for
 this orbital to the bonding ($+$) and antibonding ($-$) one
defined as
$d_{i\pm\sigma} =\frac{1}{\sqrt{2}} (d_{i1z\sigma}\pm
d_{i2z\sigma})$.
Importantly, by going beyond the approximation of projecting out
the antibonding $z^2$ orbital~\cite{Liao_PRB_2023}, we will be able to
connect the physics at the physical $N=3$ to what happens at nearby electron
fillings, including the half filling at $N=4$.
The results
thus obtained
are
 verified to be robust based on calculations using
 the rotational invariant Gutzwiller variational 
 method~\cite{LanataHellsing_PRB_2012, PengDai_CPC_2022}.

\textit{
	Global phase diagram and
	strong orbital-selectivity.~}In
Fig.~\ref{Fig:1}{\bf c} we show the evolution of the orbital-resolved
quasiparticle spectral weight $Z_\alpha$ with $U$ for $J_{\text{H}}/U=0.25$ at
electron number $N=3$ in the bilayer two-orbital Hubbard model, which
corresponds to the physical electron count
for the
 electron configuration
$d^{7.5}$ per Ni ion.
For $U\gtrsim3$ eV the system exhibits strong orbital
selectivity, with the electrons in
the $z^2$ orbitals
being more correlated than
those in the
$x^2-y^2$ orbital as 
combined effects of orbital dependent bandwidths and electron fillings
(see Figs.~S1 and S3 of
SM~\cite{SM}).

To understand the strong orbital selective
behavior at $N=3$, we examine $Z_\alpha$ with $U$ for $N=3.5$ and $N=4$.
As shown in Fig.~\ref{Fig:1}{\bf c-e},
both the quasiparticle weight renormalization and
orbital selectivity are stronger
with increasing $N$;
for $N=4$ where the model is at half filling, a Mott
transition develops with increasing $U$ to about $4$ eV. 
These results
lead to the quantitative
global phase diagram in Fig.~\ref{Fig:1}{\bf b}.
 Importantly, the system already exhibits 
the orbital-selective behavior, with the $z^2$ orbital being more correlated, at a 
crossover $U_{\rm{cr}}$ smaller than the onset of orbital-selective Mott or the 
Mott transition. This strong
orbital-selective behavior away from
half filling
can then be naturally understood
as doping a putative Mott insulator (MI) at
half filling ($N=4$), an effect
reminiscent of the strong
correlations in heavily hole-doped iron
pnictides~\cite{Yu_COSSMS_2013,Eilers_PRL_2016}. 

\begin{figure}
\includegraphics[width=0.3\textwidth]{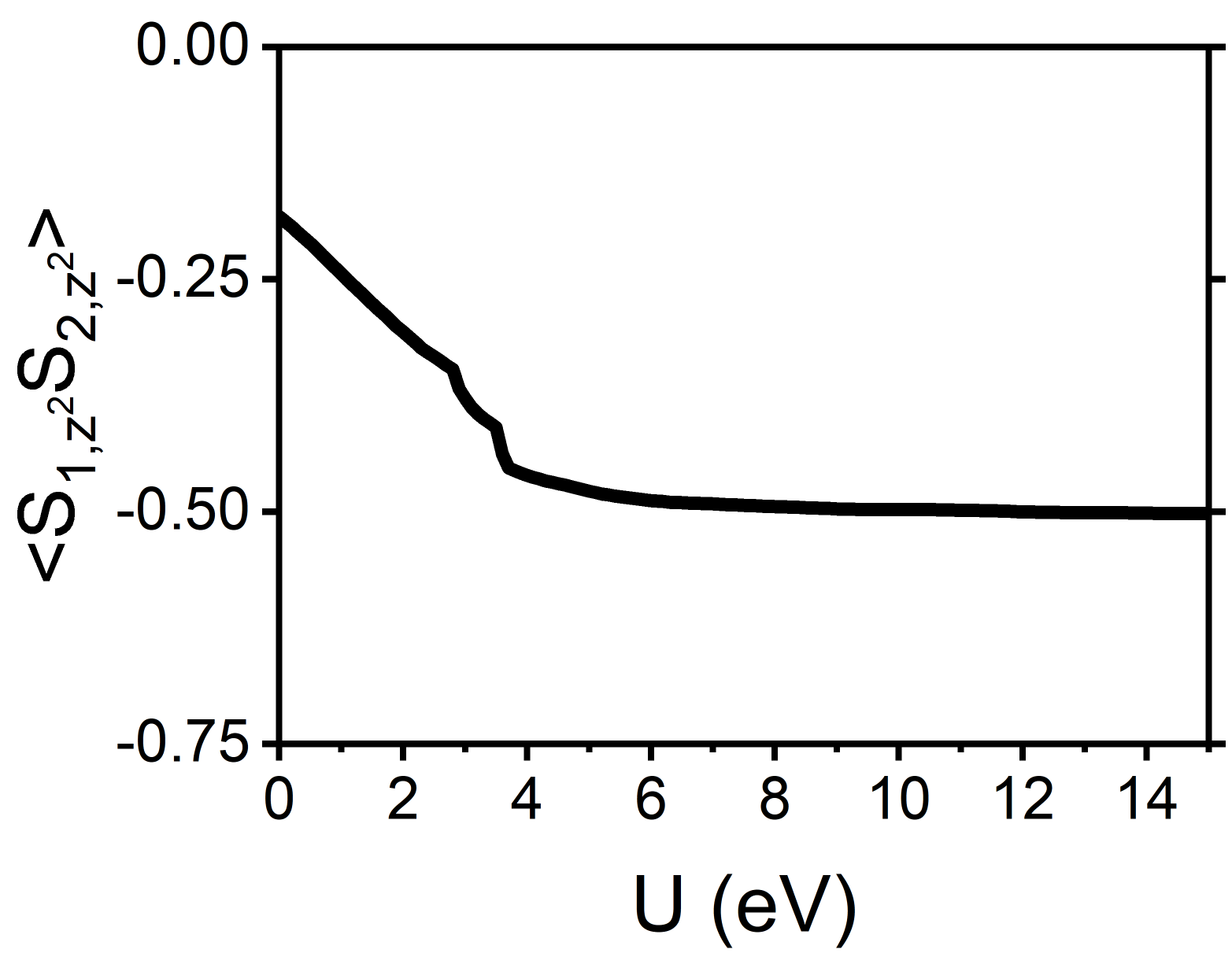}
\caption{Evolution of the interlayer
	spin correlator
in the $z^2$ orbital of the bilayer two-orbital model at $N=3$.
	 The flatness of the
	curve suggests
	interlayer spin-singlet formation
	of the $z^2$
	orbital electrons.
}
\label{Fig:2}
\end{figure}

\textit{Formation of interlayer spin singlet.~}The behavior of $Z_\alpha$
in Fig.~\ref{Fig:1}{\bf c} shows that the
strong
orbital selectivity exists in a broad regime of $U$ values.
In this regime, the two $z^2$
orbitals are nearly localized, and
are close
to half filling (Fig.~S3{\bf a} of SM~\cite{SM}).
This clearly indicates
development of quasi-localized magnetic
moments. Given the large interlayer hopping $t^{11}_{z}$ in the $z^2$ orbital,
we expect the two moments to interact via an antiferromagnetic (AFM)
superexchange
coupling and hence form an interlayer spin singlet.
We calculate the interlayer
(equal time) spin correlator
 $\langle
\mathbf{S}_{1,z^2}\cdot\mathbf{S}_{2,z^2} \rangle$ for electrons in the $z^2$
orbital, and the
result is shown in Fig.~\ref{Fig:2}.
The spin correlator decreases with increasing
$U$ monotonically and reaches
a large negative value for
$U\gtrsim 4$ eV, where the system exhibits strong orbital
selectivity. In this regime it almost saturates to about $-0.5$. The negative
sign indicates the interaction between the  moments are indeed
AFM, and the almost saturation of the correlation function implies
the formation of interlayer spin singlet. Note
that
it has a smaller amplitude than
$(-3/4)$ 
for a perfect spin singlet;
the difference
 is attributed to the influence of both the Hund's coupling 
and the deviation from half filling of the $z^2$ orbitals (see Sec.~2.1 of 
SM~\cite{SM}).

The spin singlet state has a finite gap $\Delta$
 to the
triplet excitation.
This gap can be equated to
$J_{z^2-z^2}$,
the interlayer superexchange
coupling in the $z^2$ orbital channel. This allows us to estimate the effective superexchange
coupling between these moments from
the low-temperature behavior of $\langle
\mathbf{S}_{1,z^2}\cdot\mathbf{S}_{2,z^2} \rangle$, which is detailed in
the
SM~\cite{SM}. The extracted $J_{z^2-z^2}$ values are shown in Fig.~S4{\bf b} of the
SM~\cite{SM}. The non-monotonic behavior of $J_{z^2-z^2}$ with $U$
indicates a
clear crossover between weak and strong orbital-selective electron
correlations~\cite{Ding_PRB_2019}.
While $J_{z^2-z^2}$ for the realistic 
$U\sim 6$ eV is smaller than the 
strong-coupling
expression
 $4(t^{11}_z)^2/U$ 
  of a single-orbital 
model, 
it still is sizable.

\begin{figure}
	\includegraphics[width=0.45\textwidth]{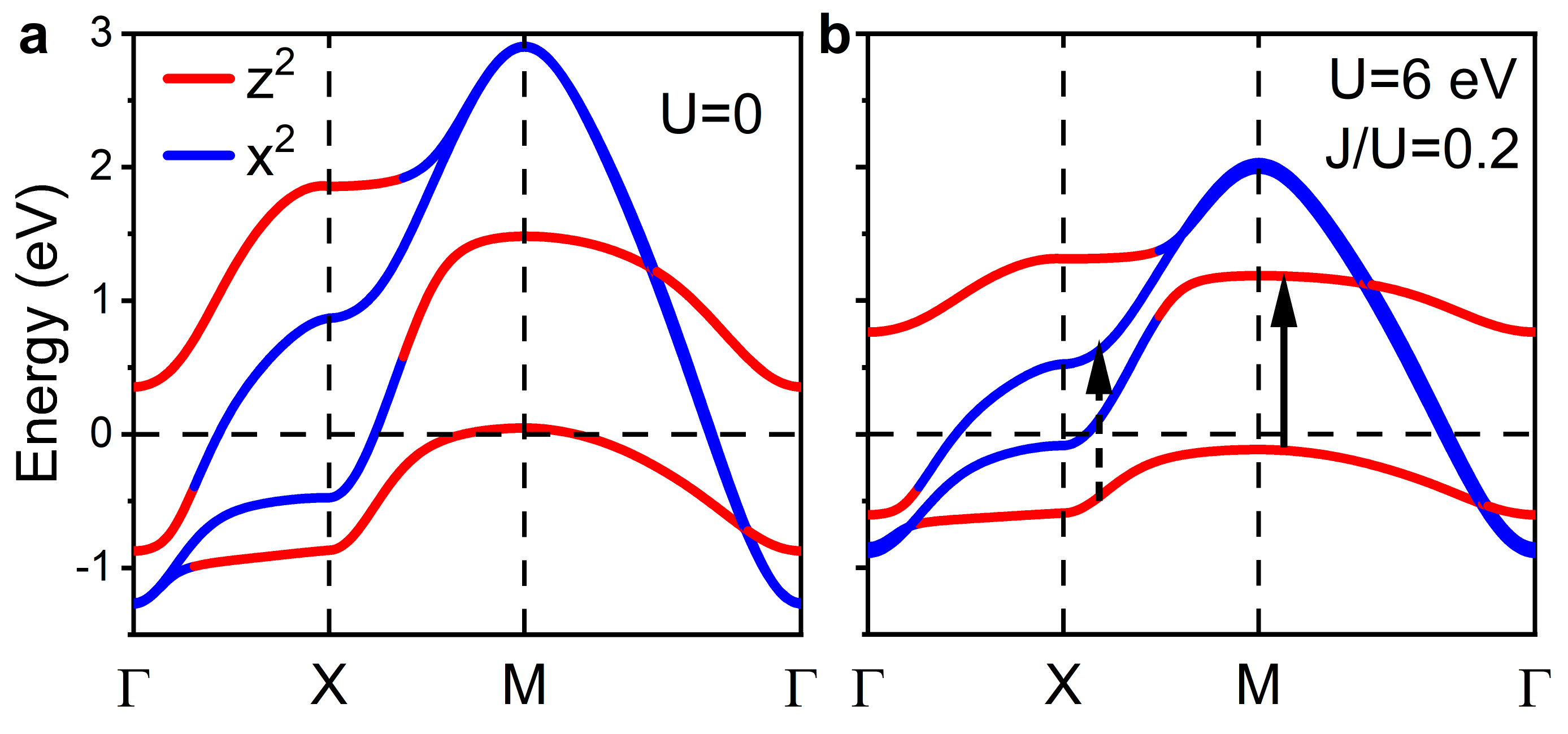}
	\caption{Calculated band structures of the bilayer two-orbital Hubbard
	model for $N=3$ at $U=0$ (in {\bf a}) and $U=6$ eV with
	$J_{\text{H}}/U=0.2$ (in {\bf
	b}). The blue and red
	colors indicate the major orbital characters of the bands are $x^2-y^2$ and
	$z^2$, respectively. The Fermi energy $E_F=0$ is set.
	The dashed and solid arrows denote typical processes that contribute to
	$\sigma_{xx}$ and $\sigma_{zz}$, respectively.}
	\label{Fig:3}
\end{figure}

\textit{Implications for spectroscopy.~}The strong orbital selectivity and the
formation of spin singlet in the model have important effects on the band
structure which can be detected by the
ARPES
spectra and
will be manifested in the optical conductivity.
In Fig.~\ref{Fig:3} we compare the band structures of the model
at $U=0$ and $U=6$ for $J_{\text{H}}/U=0.2$ at $N=3$. The correlation effect is
first seen as a substantial suppression of the overall bandwidth. Another
prominent effect is that the flat bonding $z^2$ band near the M point
sinks to
slightly below $E_F$.
This removes the small hole pocket near the M point and only
slightly changes
the volumes of
the other Fermi pockets.

To see the
orbital-selective band renormalization behavior, we extract the effective mass
enhancement factor $m^*/m_0$ of bands near $E_F$ along several cuts in the BZ
that correspond to those studied in a recent ARPES
experiment~\cite{YangZhou_NC_2024}. The cuts are
shown in Fig.~S2{\bf b}, and the $m^*/m_0$ values estimated from velocity ratios
 for $U=6$ eV are summarized
in
Table~\ref{table1}.
The bands mainly have $x^2-y^2$ orbital
characters along cuts 1 and 2,
 while
they are
dominated by the bonding ($z^2_+$)
orbital along
cuts 4 and 5.
Accordingly, the
orbital selectivity
causes an orbital-dependent
effective mass enhancement, as observed in the ARPES measurement. Along cut 3,
the band has a mixed character between the
$x^2-y^2$ and the $z^2$ orbitals and $m^*/m_0$ takes an intermediate value. In
Table~\ref{table1}, we
also list the $m^*/m_0$ values determined from the ARPES experiment~\cite{YangZhou_NC_2024}.
Our results agree with the experimental ones
semi-quantitatively.

\begin{table}
	\caption{Comparison between
	our theoretical
	results with the measurements of
	Ref.~\cite{YangZhou_NC_2024}
	on the
	effective mass enhancement factor $m^*/m_0$ of bands near $E_F$ along
	several cuts in the BZ shown in Fig.~S2 of the SM~\cite{SM}.}
	\label{table1}
	\begin{ruledtabular}
		\begin{tabular}{ccc}
			&ARPES&$U=6$ eV, $J_{\rm{H}}/U=0.25$ \\
			\hline
			Cut 1&1.8&2.5\\
			\hline
			Cut 2&2.3&1.7\\
			\hline
			Cut 3&2.6&3.3\\
			\hline
			Cut 4&7.7&4.6\\
			\hline
			Cut 5&5.0&4.6\\
		\end{tabular}
	\end{ruledtabular}
\end{table}

One surprising feature in our calculation is that the splitting between the
bonding and antibonding $z^2$ bands remains to be large at about $1$ eV
although the bandwidths are substantially renormalized under strong
correlations. This is clearly seen by comparing the bands at $U=0$ and $U=6$ eV
in Fig.~\ref{Fig:3}.
This reflects the renormalization of the electronic energy dispersion
by the sizable interlayer spin-exchange interaction  $J_{z^2-z^2}$,
as described in detail in the SM~\cite{SM}.
It can also push the bonding band to below $E_F$.
This well explains the almost temperature independence of the energy position
of the $\gamma$ band observed in ARPES because $J_{z^2-z^2}$ is much larger
than room temperature. To further verify that the exchange interaction leads to
band splitting, we perform calculation within a single-site approximation,
where the superexchange interaction is not taken into account. As shown in
Fig.~S7{\bf a}, the bonding-antibonding splitting is substantially suppressed 
in this
approximation~\cite{SM}.

\begin{figure}
	\includegraphics[width=0.45\textwidth]{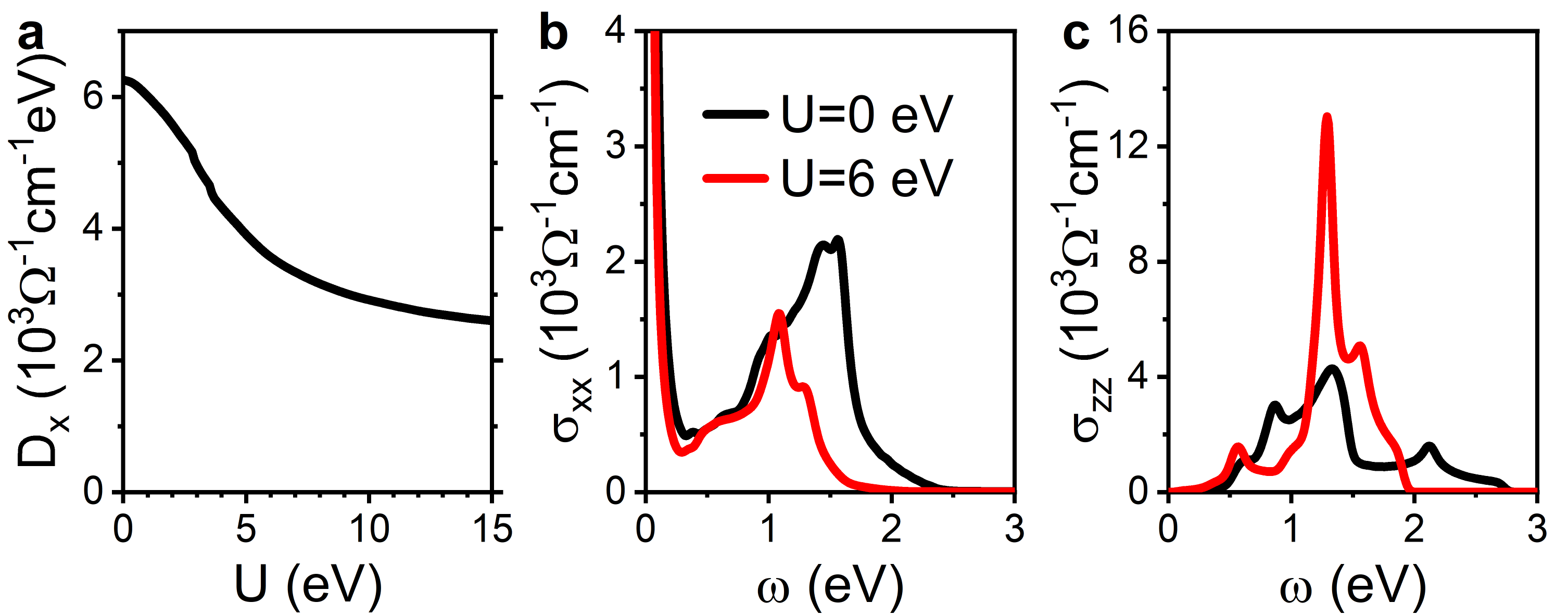}
	\caption{{\bf a} The evolution of Drude weight $D_x$ of the in-plane
	optical conductivity with $U$. {\bf b} The
	in-plane optical conductivity $\sigma_{xx}$ at $U=0$ and $U=6$ eV and
	$J_{\text{H}}/U=0.2$.  
	{\bf c} 
	The out-of-plane optical conductivity
	$\sigma_{zz}$ at $U=0$ and $U=6$ eV. A sharp peak emerges at $U=6$ eV when
	the $z^2_+$ bonding band sinks below $E_F$.}
	\label{Fig:4}
\end{figure}

We now turn to the
optical conductivity. The calculated Drude
weight $D_x$ of the in-plane component of the optical conductivity
$\sigma_{xx}$ is shown in Fig.~\ref{Fig:4}{\bf a}. $D_x$ decreases with
increasing $U$, in a way
that follows the behavior of the quasiparticle spectral
weight $Z$. We find $D_x(0)/D_x(U)$ approximately equals to the
orbital averaged $1/Z$ value, and for $U\gtrsim 6$ eV, $D_x(0)/D_x(U)\sim 2$-$3$.
This reduction factor of the Drude weight 
is sizable,
and captures the essence of the experimental observation.
Quantitatively, the ratio extracted from the optical conductivity measurement~\cite{LiuWen_NC_2024} is even larger
(on the order of  $\sim 10$, see Fig.~\ref{Fig:4}{\bf b}), which could suggest 
that the relevant interaction strength is even larger or
the need for orbital-resolved Drude weight estimate based on light polarization.

Besides the Drude weight, we also calculate the interband contribution of the
optical conductivity and the overall results for $U=0$ and $U=6$
eV with $J_{\text{H}}/U=0.2$ are shown in Fig.~\ref{Fig:4}{\bf b}.
At $U=0$, $\sigma_{xx}$
develops a broad peak at $1.3$-$1.5$ eV. With strong band
renormalization, we would expect the peak shifts drastically to low energy, as
shown in Fig.~S7{\bf b}. However, as shown in Fig.~\ref{Fig:4}{\bf b}, the 
interband
peak only slightly shifts to lower energy at about $1$ eV for $U=6$ eV.
The reason for this is revealing of the underlying physics.
The peak is mainly caused by the
transition indicated by the dashed arrow (between the bonding $z^2$ band and
the band with hybridized antibonding $z^2$ and $x^2-y^2$ orbitals) in
Fig.~\ref{Fig:3}{\bf b}, and the enhanced bonding-antibonding splitting by the
superexchange interaction of local moments keeps the transition energy to be
still at about $1$ eV.
The calculated 1eV interband peak 
is consistent with the experimental result~\cite{LiuWen_NC_2024}, corroborating 
the substantial electron correlations in the material. Note that the spectral 
weights above 2 eV observed in experiment are likely caused by transitions 
involving band at high energies, which are beyond our 
two-orbital model including Ni-$e_g$ orbitals only.

To directly probe the bonding-antibonding splitting, we propose to measure the
out-of-plane component $\sigma_{zz}$, which is contributed from intraorbital transitions by the polarization set up. As illustrated
by the solid arrow in Fig.~\ref{Fig:3}{\bf b}, when the bonding band
sinks to below
$E_F$, such an interband transition would generate a sharp peak at about $1$ eV
in $\sigma_{zz}$, as shown in Fig.~\ref{Fig:4}{\bf c}.
However, if the
bonding band crosses $E_F$, the transition associated with the portion above
$E_F$
will no longer
contribute;
$\sigma_{zz}$ would be substantially
suppressed with the peak position almost unchanged, as
can be inferred from
Fig.~\ref{Fig:4}{\bf c}. Therefore,
measuring $\sigma_{zz}$ in the
high-pressure phase
would allow
not only the determination of the bonding-antibonding splitting energy
but also the assessment about
whether the bonding state crosses the Fermi level.

\textit{Discussion and conclusion.~} The above results 
have been derived based on a 
tight-binding model for the high-pressure phase. 
We have further constructed a 
counterpart model for the low-pressure phase based on the corresponding DFT calculation,
which is given  
in Sec.~2.4 of the SM~\cite{SM}, 
leading to very similar results in the presence of
 electron correlations.

In a previous study~\cite{Liao_PRB_2023} on the correlation
effects of the bilayer two-orbital Hubbard model for La$_3$Ni$_2$O$_7$, we
further
simplified the problem by
projecting out the higher-energy antibonding $z^2$ orbital.
The calculation reported
 here is crucial, as it allows us to study the $N=3$ system
from a broader
perspective of varying $N$.
We find
the correlations in
the antibonding $z^2$ orbital
to be stronger than
its bonding counterpart at $N=3$ (see
Fig.~\ref{Fig:1}{\bf b}).
This is because,
when
 this orbital
is
allowed to be active, its high energy nature demands
a narrower bandwidth
for the kinetic energy to be minimized.
This effect causes some quantitative differences from previous study such 
as the size of the Fermi pockets and the onsite $U$ value for the OSMP.
Nonetheless, we find
robustly strong orbital selectivity,
and
we are able to understand this surprising result in the framework of
the  global phase diagram proposed here
({\it c.f.} Fig.~\ref{Fig:1}{\bf a}).

A recent experiment~\cite{Hwang_Nature_2024} on the La$_3$Ni$_2$O$_7$ thin film
showed that O$_3$ annealing
can turn the insulating sample to metallic (and becomes superconducting under
cooling), accompanied by
a valence increase of the Ni ion.
The trend of the resistivity and valence changes can be well understood within
our  proposed global phase diagram: The annealing effectively increases
the doping level, pushing the system away from the parent MI to a metal with strong
orbital-selective correlations.

An important implication of the global phase diagram is that it allows for
constructing
 a multiorbital $t$-$J$ model
to study superconductivity
that is applicable for the orbital-selective correlation regime.
Given its multiorbital nature, we expect the system to exhibit a stronger 
competition in  
pairing channels
 than in the single-orbital case: Here the	
$d$-wave pairing, favored by
the in-plane exchange
coupling between electrons in the $x^2-y^2$ orbital, 
should be in
 competition
with the extended-$s$ pairing,
driven by the interlayer exchange coupling
between electrons in the $z^2$ orbital~\cite{Duan_2025}. In the mean time, the 
orbital-selective correlation sets the stage for discussing the stability of 
superconductivity and competing electronic (either spin or charge density wave) 
orders, which is crucial in understanding the pressure driven high-temperature 
superconductivity in multi-layer nickelates.

In summary, we
have studied
the orbital-dependent electron correlations in a bilayer
two-orbital Hubbard model for La$_3$Ni$_2$O$_7$.
We find strong orbital selectivity away from half filling that is
ultimately
 anchored by
a Mott insulator at half filling,
a result that is encoded in a global phase diagram  
proposed here
(Fig.~\ref{Fig:1}{\bf a}).
At the physical electron count $N=3$, we show that the strong orbital
selectivity leads to the formation of interlayer singlets between electrons in the $z^2$
orbitals. The latter
contribute  to the strong renormalization of the band
structure, in particular an enhanced splitting between the bonding and
antibonding $z^2$ bands.
Together, these effects explain all four key
features
observed in the ARPES and optical conductivity
measurements that
we outlined in the introduction. Given the importance of
both the electronic structure and electron correlations to
superconductivity, our results provide a basis to understand
not only the normal state but also the high temperature superconductivity
of the bilayer nickelates.
More generally, our work enriches the microscopic physics that is crucial
to correlated superconductivity in general.

%
%

%

\begin{acknowledgments}
We thank X. Dai, W. Ding, K.-S. Lin, S. Peng, and H. Y. 352
Hwang for useful discussions. This work has been supported 353
in part by the National Natural Science Foundation of China 354
(Grants No. 12334008 and No. 12174441). R.Y. acknowl- 355
edges the Nickel-based High temperature Superconductivity 356
Workshop II at Westlake University where part of this work 357
was discussed. Work at Rice was primarily supported by the 358
U.S. Department of Energy, Office of Science, Basic Energy 359
Sciences, under Award No. DE-SC0018197. Q.S. and L.C. 360
acknowledge the hospitality of the Aspen Center for Physics, 361
which is supported by NSF Grant No. PHY-2210452 and a 362
grant from the Simons Foundation (1161654, Troyer). 
\end{acknowledgments}


%


\clearpage

\onecolumngrid

	\begin{center}
			\textbf{\large SUPPLEMENTAL MATERIAL --
					Orbital-selective electron correlations in
					high-$T_{\rm c}$\\bilayer nickelates:
					from a global phase diagram
					to implications for spectroscopy}\\[.2cm]
		\end{center}



\setcounter{equation}{0}
\setcounter{page}{1}

\setcounter{figure}{0}
\setcounter{table}{0}
\makeatletter
\renewcommand{\thefigure}{S\@arabic\c@figure}
\renewcommand{\thetable}{S\@arabic\c@table}

\renewcommand{\theequation}{S\@arabic\c@equation}


\onecolumngrid


\section{S1 Details of the tight-binding model}

To obtain the tight-binding model, we first perform the density functional 
theory (DFT) calculations under the plane wave basis using projector 
augmented-wave potentials with the Perdew-Burke-Ernzerhof 
exchange-correlation functional as implemented in the Quantum Espresso (QE) 
code~[74]. 
The energy cutoff for the plane wave basis is set to 600 eV.
The k-mesh for sampling the Brillouin zone is 
$6\times6\times6$ 
with respect to a symmetric unit cell for the high-pressure phase.
The convergence threshold for self-consistent calculations is 
taken as
$10^{-6}$ eV.
When performing the crystal structure optimization, we keep the 
experimentally measured lattice constants unchanged and only relax the 
atomic positions within the unit cell. The convergence threshold is 
that all forces are smaller than $10^{-3}$ eV/$\AA$.

\begin{table}[b!]
	\caption{Tight-binding parameters (in eV) of the bilayer two-orbital 
		model in this work.
		$\epsilon_{\alpha}$ denotes the onsite-energy of the $\alpha$ 
		orbital.
		$\alpha=1$, $2$ represent $z^2$ and $x^2-y^2$ ($x^2$) orbitals, 
		respectively. We have kept hopping parameters up to the sixth 
		nearest-neighbors. $\mu$ is the index of neighbors, for example, 
		$\mu=x$ and $\mu=xy$ 
		refer to the nearest-neighbor (n.n.) and next 
		nearest-neighbor (n.n.n.) in-plane hoppings, $\mu=z$ and $\mu=xz$ 
		refer to the n.n. and n.n.n. interlayer hoppings, respectively.
		$\pm$ means the parameters are positive along the x direction and 
		negative along the y direction.
	}
	\label{tables1}
	\begin{ruledtabular}
		\begin{tabular}{cccccccc}
			&$\alpha=1$&$\alpha=2$&&&&&\\
			\hline
			$\epsilon_{\alpha}$&10.5124&10.8716&&&&&\\
			\hline
			\hline
			$t^{\alpha\beta}_{\mu}$&$\mu=x$&$\mu=xy$&$\mu=xx$&$\mu=xxy$&$\mu=xxx$&$\mu=z$&$\mu=xz$\\
			\hline
			$\alpha\beta=11$&-0.1123&-0.0142&-0.0181&-0.0029&-0.0084&-0.6420&0.0257\\
			\hline
			$\alpha\beta=22$&-0.4897&0.0686&-0.0605&-0.0078&-0.0256&0.0029&0.0006\\
			\hline
			$\alpha\beta=12$&$\pm$0.2425&&$\pm0.0322$&$\pm0.0038$&$\pm0.0134$&&$\mp$0.0370\\
		\end{tabular}
	\end{ruledtabular}
			%
\end{table}

\begin{figure}[h!]
	\includegraphics[width=0.6\textwidth]{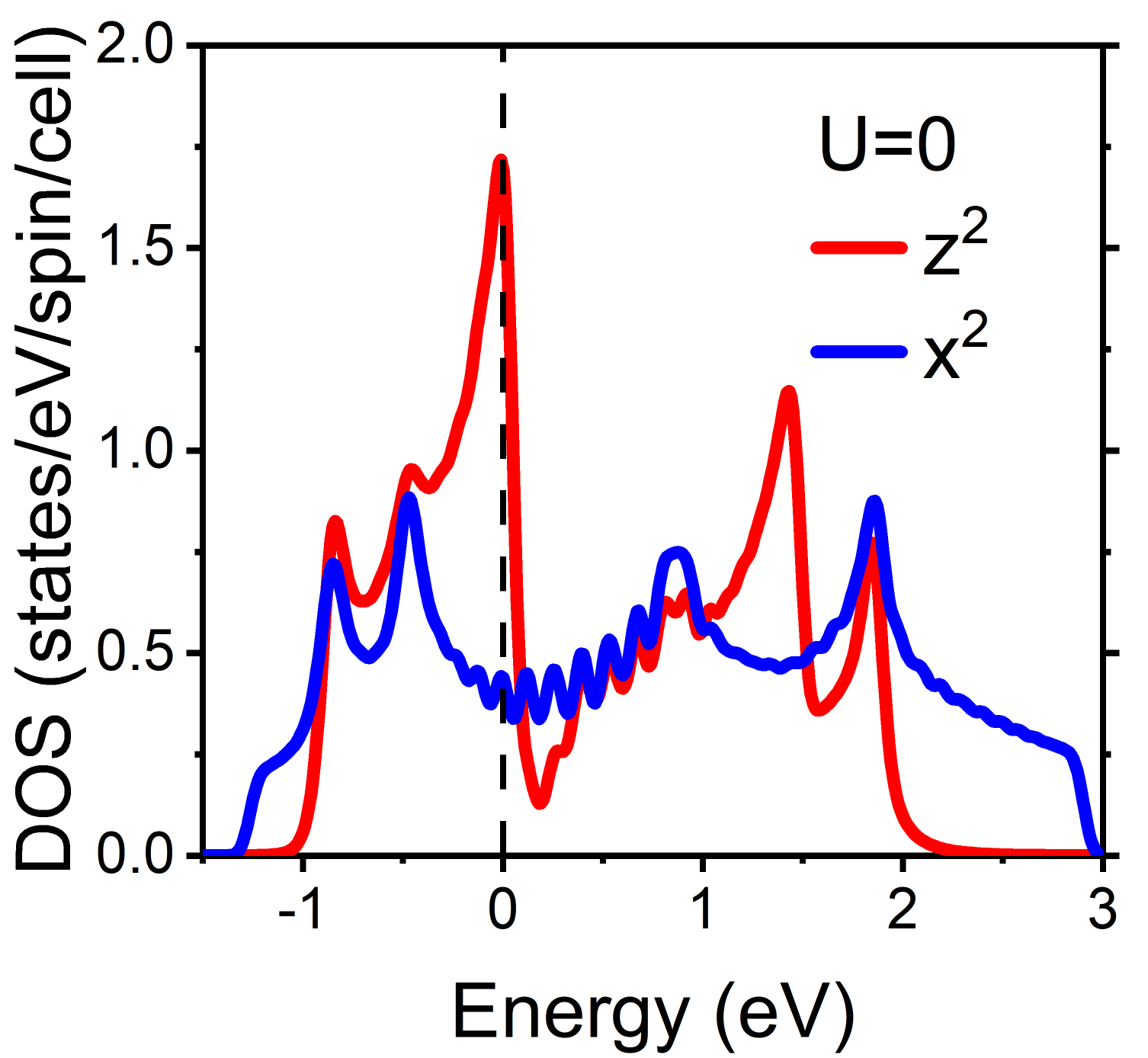}
	\caption{Density of states (DoS) projected to the $x^2-y^2$ ($x^2$) and 
	$z^2$ orbitals of the bilayer two-orbital tight-binding model
		({\it i.e.}, for
		$U=0$ in the Hubbard model).}
	\label{s1}
\end{figure}

Based on the DFT results, the tight-binding models are constructed with Ni e$_g$ 
orbitals using the maximally localized Wannier functions scheme as implemented 
in the WANNIER90 code~[75]. 
The model contains hopping parameters up to the 
6th nearest-neighbors and the tight-binding parameters are listed in 
Table~\ref{tables1}.

\begin{figure}
	\includegraphics[width=0.6\textwidth]{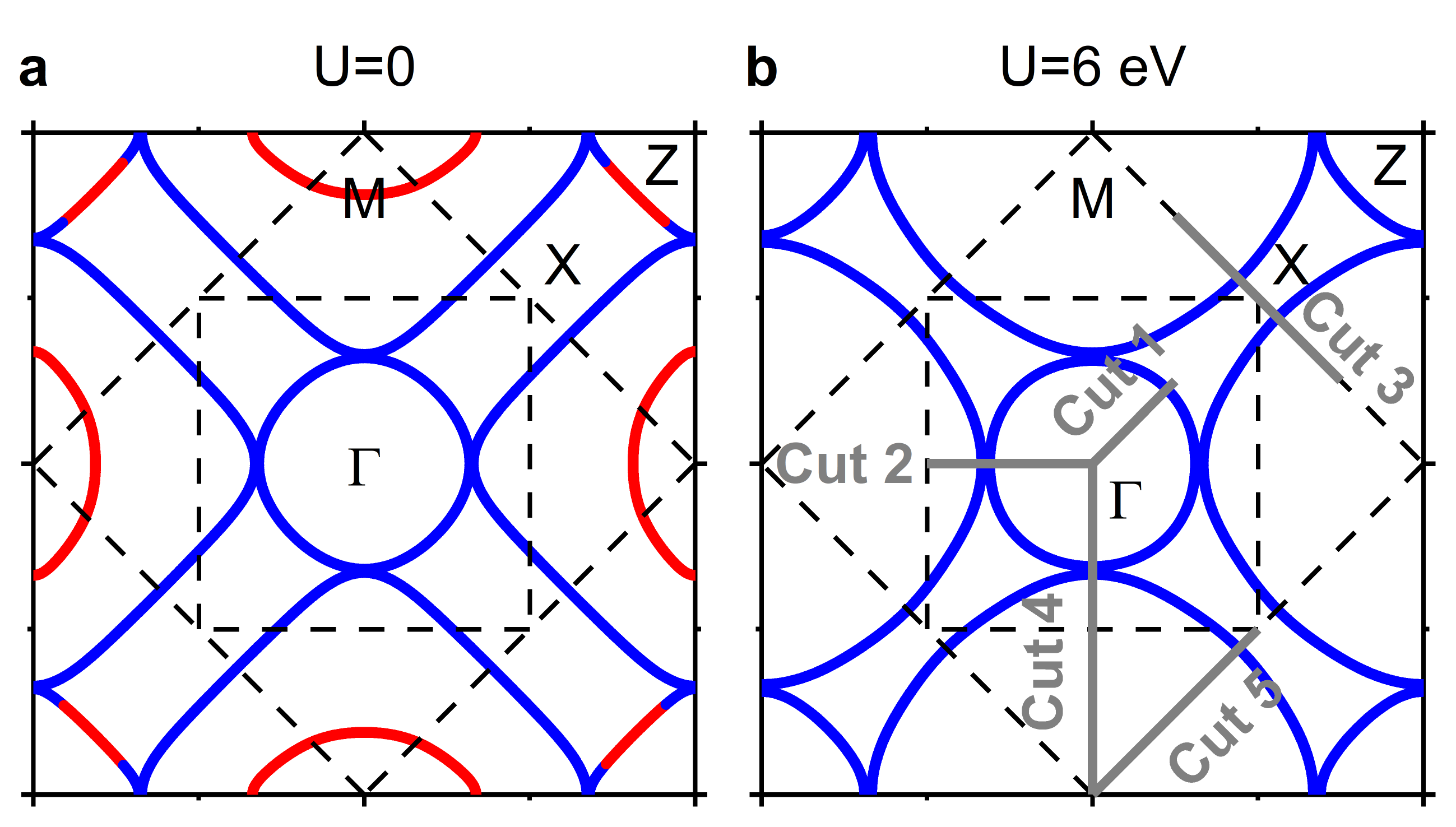}
	\caption{Fermi surface of the bilayer two-orbital Hubbard model at $U=0$ (in 
	{\bf a}) and $U=6$ eV and $J_{\rm{H}}/U=0.2$ (in {\bf b}).
		The blue and red
		colors indicate that the dominant orbital characters are $x^2-y^2$ and 
		$z^2$, respectively.
		The minority orbital component, which is especially substantial near X 
		in the Brillouin zone, is not marked.}
	\label{s2}
\end{figure}

The band structure of the tight-binding model is shown in Fig.~3{\bf a} of 
the main text. It corresponds to the band structure in the high-pressure 
phase. The band structure in the
low-pressure phase can be obtained by 
folding the bands to the Brillouin zone corresponding to a double-in-size 
primitive cell.
To show this model can indeed describe the bandstructure in the 
low-pressure phase, we constructed another tight-binding model for the 
low-pressure phase based on DFT bandstructure at ambient pressure. Details 
on the comparison of bandstructures between the low-pressure and 
high-pressure models are given in Sec.~2.4.
In Fig.~3{\bf a} we find the splitting between the bonding and antibonding $z^2$ 
bands
to be  $2|t_z^{11}|\approx1.3$ eV. The bonding band crosses the Fermi level (at 
$E_F=0$), causing a small hole pocket centered at M point in the Brillouin zone 
as shown in Fig.~S2{\bf a}. Besides this pocket, the Fermi surface also contains 
an electron pocket centered at $\Gamma$ point and a large hole pocket, both
of which are
dominated
by the $x^2-y^2$ orbitals. The density of states projected on to the two $e_g$ 
orbitals is shown in Fig.~S1. Compared to the $x^2-y^2$ orbital, the $z^2$ 
orbital has a much narrower bandwidth. We then expect stronger correlation 
effects in this orbital, as shown in the main text.


\begin{figure}
	\includegraphics[width=0.7\textwidth]{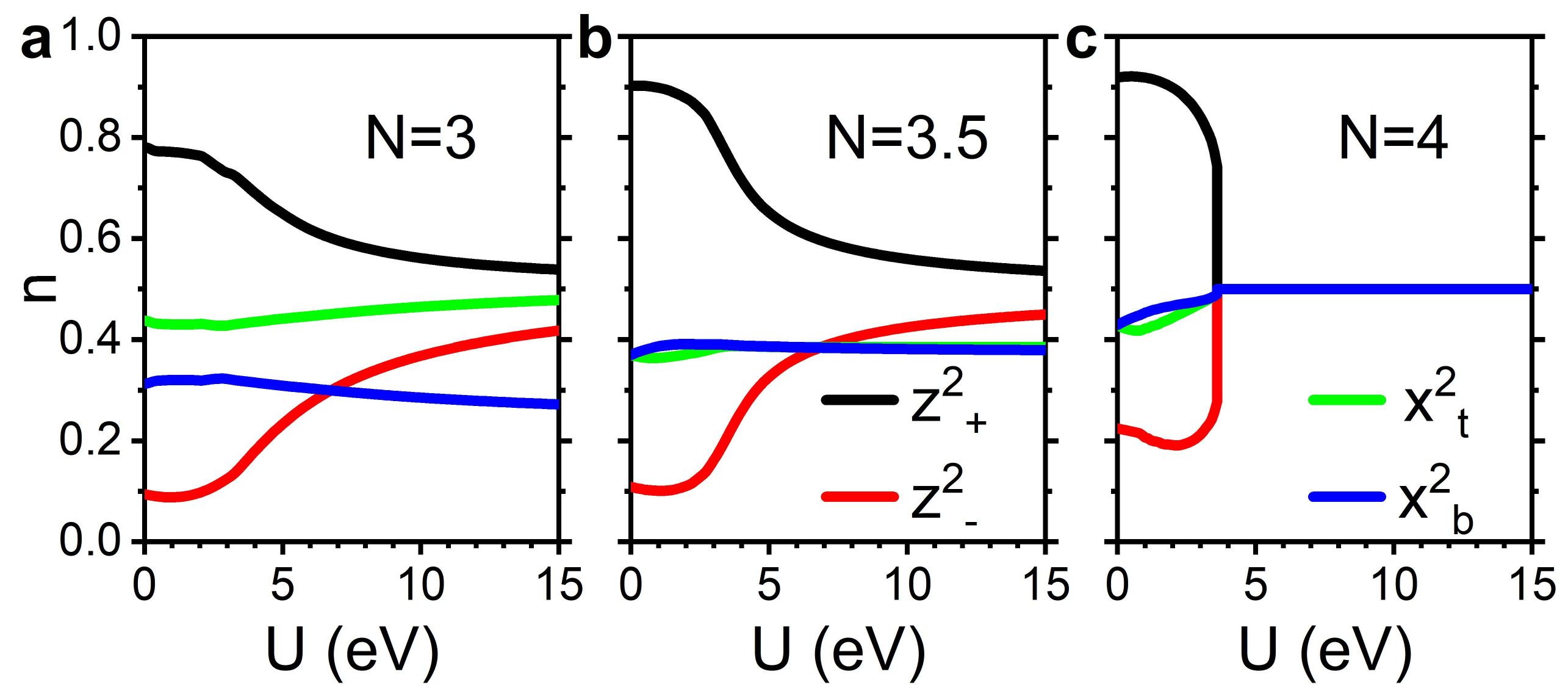}
	\caption{{\bf a}-{\bf c} Evolution of the electron filling in each 
		orbital with $U$ for $J_{\rm{H}}/U=0.25$ of the bilayer two-orbital 
		model at $N=3$, $N=3.5$, and $N=4$, respectively.}
	\label{s3}
\end{figure}

\section{S2 Supplemental results on orbital-selective electron correlation 
	effects}
\subsection{S2.1 Orbital selectivity and formation of interlayer
	$z^2$ spin singlet}
We have shown in the main text that the system contains strong orbital-selective 
electron correlations at the the physical electron count $N=3$, and this is 
understood
in terms of a proximity to an orbital-selective Mott phase, which in turn is near
a Mott insulator at $N=4$. 
In Fig.~\ref{s3}{\bf a}-{\bf c} we display the orbital-resolved electron 
fillings corresponding to $N=3$, $N=3.5$, and $N=4$. Away from half-filling 
and at large $U$ limit, the electrons in the $z^2$ orbitals are closer to 
half-filling then in the $x^2-y^2$ orbitals. This, together with the 
narrower bandwidth projected to the $z^2$ orbital, accounts for the strong 
orbital-selective correlations away from half-filling.

\begin{figure*}
	\includegraphics[width=0.7\textwidth]{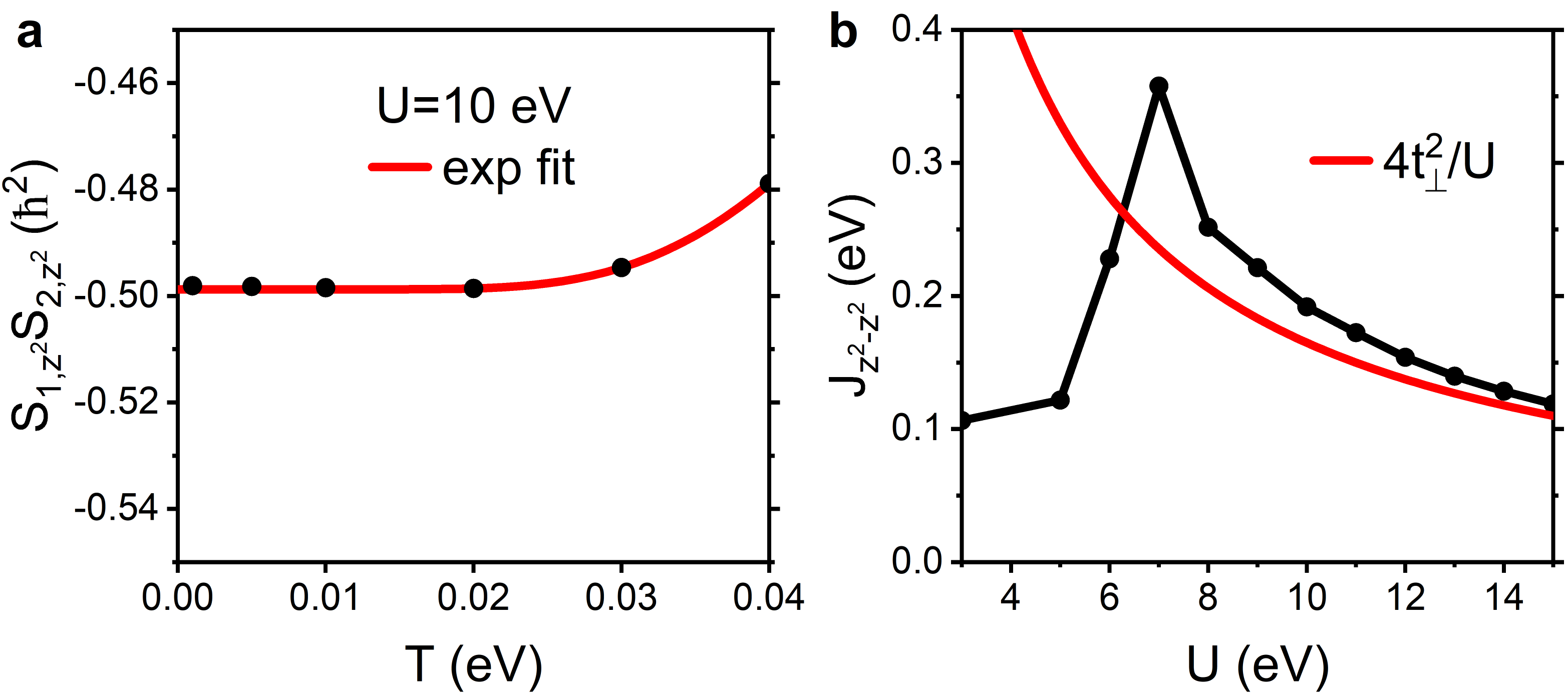}
	\caption{{\bf a} Finite-temperature spin-spin correlation function between 
	electrons of the $z^2$ orbital in the top and bottom layers. The red line is 
	a fit using the exponential function in Eq.~\eqref{Eq:Spingap}. {\bf b} 
	Extracted AFM superexchange coupling $J_{z^2-z^2}$ with $U$ for 
	$J_{\rm{H}}/U=0.2$. The red line shows a function $4(t^{11}_z)^2/U$.}
	\label{s4}
\end{figure*}

With the much stronger 
correlations in $z^2$ orbitals,
the results suggest that quasi-localized $S=1/2$ moments develop in these 
orbitals, as discussed in the main text. The claculated spin-spin 
correlation function in Fig.~2 of the main text implies that these 
quasi-local moments form interlayer spin singlets under antiferromagnetic 
(AFM) superexchange interaction. To extract the exchange coupling 
$J_{z^2-z^2}$, we calculate the spin-spin correlation function at low 
temperatures. When two moments form a spin singlet, it is gapped to the 
triplet excitations, and this causes the spin-spin correlation function to 
have an exponential behavior with the temperature, {\it i.e.},
\begin{equation}\label{Eq:Spingap}
	\langle \mathbf{S}_{1,z^2}\cdot \mathbf{S}_{2,z^2} \rangle \sim a + b 
	e^{-J_{z^2-z^2}/T}.
\end{equation}
Note that the excitation gap equals to the exchange coupling $J_{z^2-z^2}$, and 
considering that the moments are quasi-localized, we leave coefficients $a$ and 
$b$ in Eq.~\eqref{Eq:Spingap} as free parameters. We fit the finite-temperature 
spin-spin correlation data with Eq.~\eqref{Eq:Spingap}, and a typical fitting 
result is shown in Fig.~\ref{s4}{\bf a}. We can then extract the effective 
superexchange coupling $J_{z^2-z^2}$. The extracted values with $U$ for 
$J_{\rm{H}}/U=0.2$ is shown in Fig.~\ref{s4}{\bf b}. In the
large $U$ limit, $J_{z^2-z^2}$ increases with decreasing $U$ and approximately
follows the standard
atomic limit behavior $J\sim 4(t^{11}_{z})^2/U$ in a single-orbital model.
In general, the Hund's coupling in a multiorbital model should give a
correction to the $J$ value in the single-orbital
case~[72]. 
Here we find $J_{z^2-z^2}$ behaves
similarly to $J$ in
the single-orbital limit because the $x^2-y^2$ orbital is much less correlated
and further away from half-filling compared to the $z^2$ orbital. 
However, the Hund's coupling still affects the saturation value of the 
spin-spin correlation function as discussed below. $J_{z^2-z^2}$
develops a peak at $U\sim7$ eV and decreases rapidly with decreasing $U$.
Together with the fast suppression of the magnitude of $\langle
\mathbf{S}_{1,z^2}\cdot\mathbf{S}_{2,z^2} \rangle$, this leads us to the
conclusion that well defined quasilocalized moments are not yet developed in
the small $U$ regime ($U\lesssim4$ eV). Note that
the orbital selectivity
is weak
in this
part of the phase diagram.

One significant difference to the single-orbital model is that the 
spin-spin 
correlation function $C=\langle \mathbf{S}_{1,z^2}\cdot 
\mathbf{S}_{2,z^2}\rangle$ saturates to $C=-1/2$ instead of $C=-3/4$ in the 
large $U$ and zero-temperature limit. 
Note that, in the two-orbital model for 
La$_3$Ni$_2$O$_7$, one has to consider the effects of the physical filling at 
$N=3$ 
and the finite Hund's coupling $J_{\rm{H}}$. Even in the limit with the $z^2$ 
orbital 
being at half filling, there is one electron in either the top or bottom layer 
$x^2-y^2$ orbital. Under the finite Hund's coupling, this electron tends to form 
a 
spin triplet with the electron in the $z^2$ orbital at the same site. This 
effect competes with the spin singlet tendency of electrons in the $z^2$ 
orbital and suppresses the spin-spin correlation function from the ideal value 
$C=-3/4$ to $C=-1/2$.

\begin{figure}
	\includegraphics[width=0.6\textwidth]{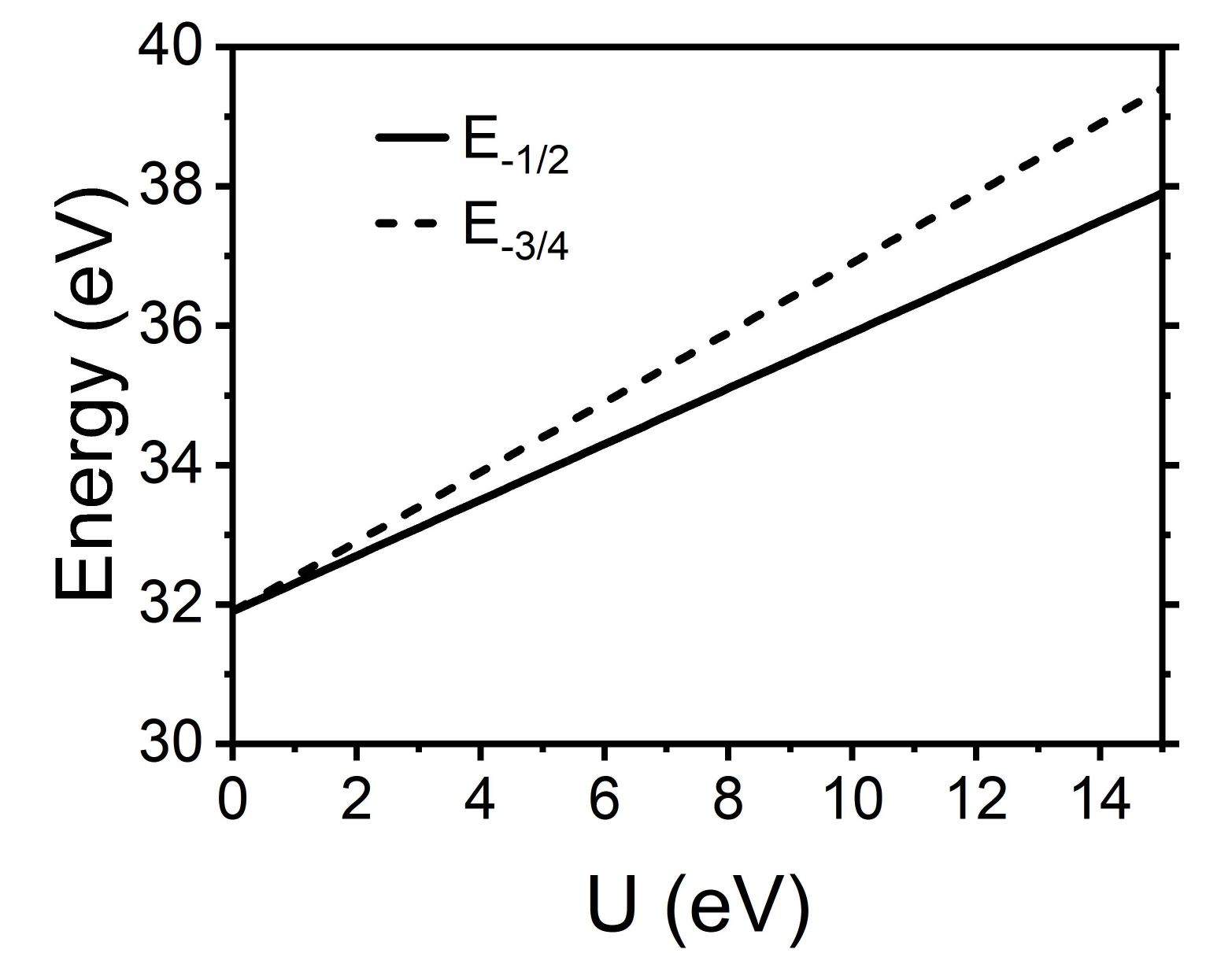}
	\caption{Comparison of energies of the states $|C=-1/2\rangle$ (solid 
		curve) and $|C=-3/4\rangle$ (dashed curve) associated with spin-spin 
		correlation functions $C=-1/2$ and $C=-3/4$, respectively.}
	\label{s5}
\end{figure}

To show this clearly, we compare the energies of two states with the spin-spin 
correlation function $C=-1/2$ and $C=-3/4$, respectively. The two states are 
denoted as $|C=-1/2\rangle$ and $|C=-3/4\rangle$, respectively:
\begin{eqnarray}
	\begin{aligned}
		|C=-1/2\rangle=
		&\frac{1}{\sqrt{3}}(d^{\dagger}_{2,z^2\uparrow} 
		d^{\dagger}_{1,z^2\downarrow} d^{\dagger}_{1,x^2\downarrow} 
		-d^{\dagger}_{1,z^2\uparrow} d^{\dagger}_{2,z^2\downarrow} 
		d^{\dagger}_{2,x^2\downarrow})|0\rangle\\
		-&\frac{1}{\sqrt{12}}(d^{\dagger}_{2,z^2\uparrow} 
		d^{\dagger}_{1,z^2\downarrow} d^{\dagger}_{2,x^2\downarrow} 
		+d^{\dagger}_{1,z^2\uparrow} d^{\dagger}_{1,x^2\downarrow} 
		d^{\dagger}_{2,z^2\downarrow})|0\rangle\\
		+&\frac{1}{\sqrt{12}}(d^{\dagger}_{2,x^2\uparrow} 
		d^{\dagger}_{1,z^2\downarrow} d^{\dagger}_{2,z^2\downarrow} 
		+d^{\dagger}_{1,x^2\uparrow} d^{\dagger}_{1,z^2\downarrow} 
		d^{\dagger}_{2,z^2\downarrow})|0\rangle,
	\end{aligned}
\end{eqnarray}
and one of the degenerate $|C=-3/4\rangle$ is
\begin{eqnarray}
	\begin{aligned}
		|C=-3/4\rangle= \frac{1}{2}(d^{\dagger}_{2,z^2\uparrow} 
		d^{\dagger}_{1,z^2\downarrow} +d^{\dagger}_{1,z^2\uparrow} 
		d^{\dagger}_{2,z^2\downarrow}) (d^{\dagger}_{1,x^2\downarrow} 
		-d^{\dagger}_{2,x^2\downarrow})|0\rangle.
	\end{aligned}
\end{eqnarray}
The corresponding energies of these two states are plotted in Fig.~\ref{s5}.
One sees that the $|C=-1/2\rangle$ state has a lower energy than the 
$|C=-3/4\rangle$ one, and the energy difference increases with increasing $U$. 
This indicates that the $|C=-1/2\rangle$ state is stabilized over the 
$|C=-3/4\rangle$ one by the Hund's coupling $J_{\rm{H}}$ 
for the fixed ratio
$J_{\rm{H}}/U$.

\begin{figure}
	\includegraphics[width=0.6\textwidth]{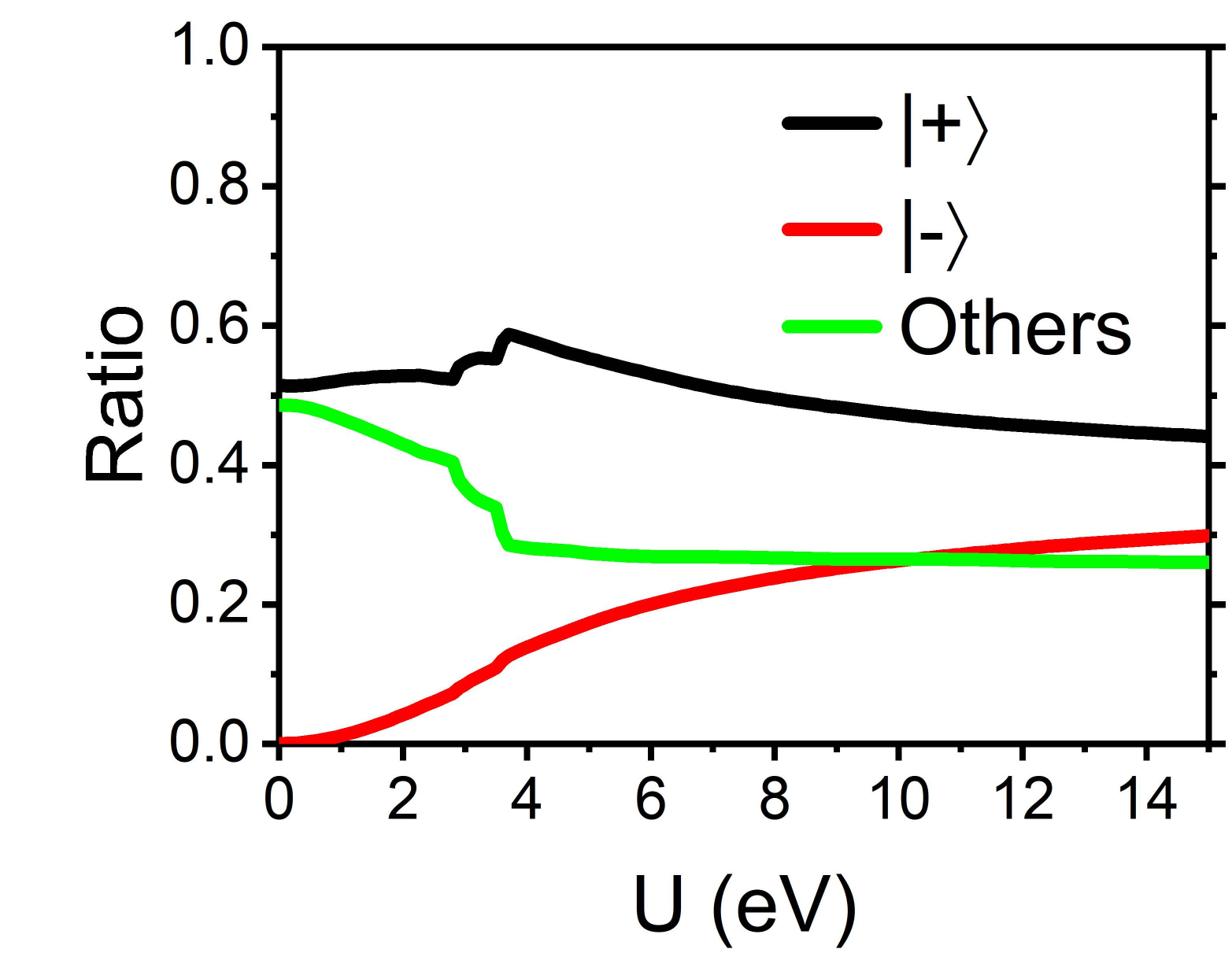}
	\caption{Evolution of ratios of $|+\rangle$, $|-\rangle$, and other 
		configurations in the ground-state wave function with $U$.}
	\label{s6}
\end{figure}

We can further estimate the spin-spin correlation function $C$ as follows. 
In 
the large $U$ limit ($U\gtrsim 6$ eV), the ground-state wave 
function projected to the bonding and antibonding $z^2$ orbital basis can be 
written as 
\begin{equation}
	|\psi \rangle \approx p_+ |+\rangle - p_- |-\rangle + \ldots
\end{equation}
where 
$|\pm\rangle=d^{\dagger}_{z2\pm\uparrow}d^{\dagger}_{z2\pm\downarrow}|0\rangle$,
$|0\rangle$ refers to the fermion vacuum, and the coefficients $p_{\pm}>0$.
The electron fillings of the bonding and antibonding orbitals are
\begin{equation}\label{Eq:filling}
	n_\pm = \langle \psi | \hat{n}_{\pm} | \psi \rangle /2 \approx p^2_{\pm},
\end{equation}
where $\hat{n}_{\pm} = \sum_\sigma d^{\dagger}_{\pm\sigma}d_{\pm\sigma}$.
In Fig.~\ref{s6} we plot the evolution of the ratios of $|\pm\rangle$ and other 
configurations with $U$. One sees that the behaviors of $p^2_{\pm}$ resemble 
those of $n_{\pm}$ in Fig.~S3. The small 
difference comes from configurations other than $|\pm\rangle$, whose ratio is 
almost independent of $U$ for $U\gtrsim 6$ eV.
Another observation from Fig.~\ref{s6} is that we can approximate 
$p^2_{\pm}\approx \bar{P} \pm x$, where the deviation $x\ll \bar{P}$ and 
$x\rightarrow0$ when 
$U\rightarrow\infty$.

The spin-spin correlation function 
\begin{equation}\label{Eq:SpinCorr1}
	C=\langle \mathbf{S}_{1,z^2}\cdot \mathbf{S}_{2,z^2}\rangle \approx 
	\frac{1}{4} - \langle \psi|s\rangle\langle s|\psi\rangle, 
\end{equation}   
where $|s\rangle = 
\frac{1}{\sqrt{2}} (d^{\dagger}_{1\downarrow} d^{\dagger}_{2\uparrow} 
- d^{\dagger}_{1\uparrow} d^{\dagger}_{2\downarrow})|0\rangle$ is the 
inter-layer spin singlet in the $z^2$ orbital. Note that $|s\rangle = 
\frac{1}{\sqrt{2}} (|+\rangle - |-\rangle)$, we then have 
\begin{eqnarray} \label{Eq:SpinCorr2}
	C &\approx& 
	\frac{1}{4} [1-2(p_+ + p_-)^2],\nonumber\\
	&\approx& (\frac{1}{4} - 2\bar{P}) + \frac{x^2}{2\bar{P}}.
\end{eqnarray}
From Fig.~\ref{s6}, we find $\bar{P}\approx0.37$, so that $C$ 
saturates to $1/4 - 2\bar{P} 
\approx -1/2$. Directly from Eqns.~\eqref{Eq:SpinCorr1} and 
\eqref{Eq:SpinCorr2}, we see that the deviation of $C$ from the ideal value 
$-3/4$ comes from configurations other than singlet, which reduces $2\bar{P}$ 
from the ideal value $1$ to about $0.75$.

As another interesting observation, we find the saturation process of the 
spin-spin correlation function $C$ to be much faster than that of the electron 
fillings, as one can clearly see by comparing Fig.~2 of the main text and 
Fig.~\ref{s3}. The reason is as follows: From Eq.~\eqref{Eq:filling}, the 
electron filling $n_\pm\approx \bar{P}\pm x$. The saturation of $n_{\pm}$ is 
proportional to $x$. On the other hand, from Eq.~\eqref{Eq:SpinCorr2}, 
$C\approx (1/4 - 
2\bar{P}) + \frac{x^2}{2\bar{P}}$, the saturation of $C$ is proportional to 
$x^2$, which is a higher order correction as $U\rightarrow\infty$. 

\subsection{2.2 Additional bonding-antibonding splitting induced by the 
	superexchange interaction}

As shown in Fig.~3{\bf b}, the splitting between the bonding and antibonding 
$z^2$ bands is still sizable under strong electron correlations at $U=6$ eV. 
This is because the AFM superexchange interaction between quasi-localized 
moments in the $z^2$ orbital can cause an additional splitting between the 
bonding and antibonding bands.
To understand this, recall that the strong
orbital selectivity in the system causes formation of the interlayer spin
singlet.
The Hamiltonian
describing the interacting moments is
\begin{equation}
	H_{\text{SE}} = J_{z^2-z^2} \mathbf{S}_{1,z^2}\cdot\mathbf{S}_{2,z^2}.
	\label{eq-SE}
\end{equation}
Given the electrons in the $z^2$ orbital are not fully localized,
this term affects the bandstructure. To see this, we
rewrite $H_{\text{SE}}$ in the spin singlet channel
\begin{eqnarray}
		H_{\rm{SE}} &=& -\frac{J_{z^2-z^2}}{2} D_{12}^\dag D_{12},
\end{eqnarray}
where $D_{12}=\sum_\sigma d_{1\sigma}^\dag d_{2\sigma}$ is the singlet operator
in the $z^2$ orbital.
In the bonding-antibonding basis, $D_{12}=n_+ - n_-$,
where $n_{\pm} = \sum_\sigma d_{\pm\sigma}^\dag d_{\pm\sigma}$. After a
saddle-point
decomposition we get
\begin{eqnarray}\label{Eq:split}
		H_{\rm{SE}} &\approx& -J_{z^2-z^2} (\langle n_+\rangle - \langle 
		n_-\rangle)
		(n_+ - n_-) \, .
\end{eqnarray}
Here,
$\langle n_+\rangle - \langle n_-\rangle$
is
precisely the difference of electron filling between the bonding and antibonding
states. According to Eq.~\eqref{Eq:split}, the interacting moments
cause an
additional splitting between the bonding and antibonding bands. For
intermediate $U$ values this splitting can be sizable as shown in
Fig.~3{\bf b} of the main text.

\subsection{2.3 Bandstructure and optical conductivity
	calculated from the single-site approximation}

\begin{figure}
	\includegraphics[width=0.7\textwidth]{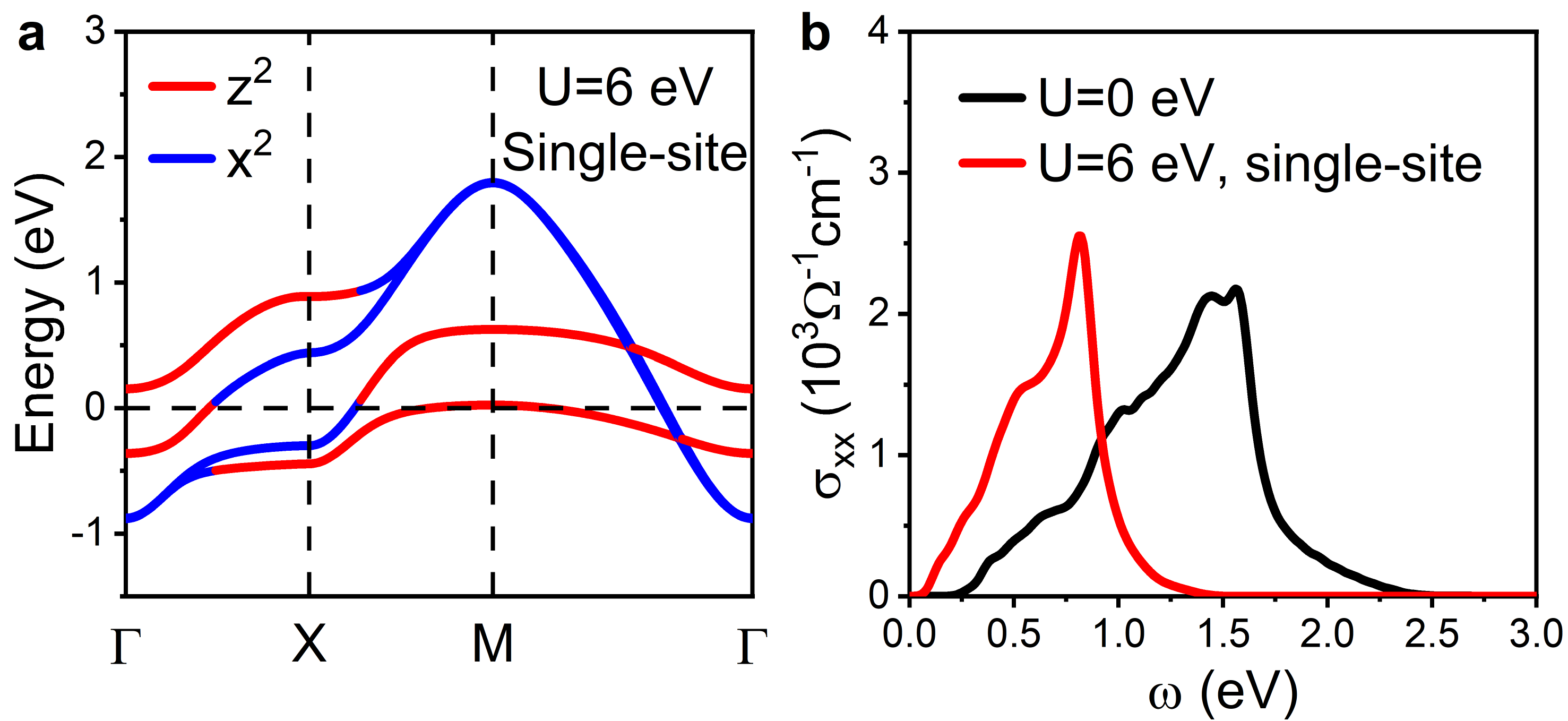}
	\caption{{\bf a} Bandstructures calculated at $U=6$ eV and $J_{\rm{H}}/U=0.2$
		within the single-site approximation (which does not consider any 
		interlayer AFM superexchange interaction between the quasi-localized 
		moments). {\bf b} Optical conductivity
		within the single-site approximation at $U=0$ and $U=6$ eV with 
		$J_{\rm{H}}/U=0.2$.
	}
	\label{s7}
\end{figure}

In this section, we show the
results of bandstructure and optical conductivity within a single-site 
approximation and
compare
the results with those calculated based on the two-site unit cell and discussed 
in the main text.
Note that within the singlet-site approximation, the interlayer correlation is 
only taken into account at a Hatree-Fock-like mean-field level so that the 
effective AFM interaction between quasi-local moments due to the singlet 
formation is not included. Therefore, the comparison between the single- and 
two-site calculations can clarify the effects of the interlayer spin singlet 
formation.

We first show the calculated bandstructure within the single-site 
approximation at $U=6$ eV and $J_{\rm{H}}/U=0.2$ in Fig.~\ref{s7}{\bf a}. 
Compared to the $U=0$ bandstructure in Fig.~3{\bf a} of the main text, 
there is a prominent reduction of the overall bandwidth as a correlation 
effect. Because the interlayer hopping $t^{11}_z$ is also renormalized and 
the splitting between the bonding and antibonding $z^2$ bands is
$2t^{11}_z$, the splitting between these two bands is substantially suppressed 
from about $1.3$ eV to about $0.5$ eV.
This is to be contrast
with the calculated result
based on a two-site unit cell, where the separation between the two $z^2$ bands
remains to be about $1$ eV, as shown in Fig.~3{\bf b} of the main text.
Such a big difference is because the AFM superexchange interaction between the 
quasi-local moments has an additional contribution to the bonding-antibonding 
splitting, as
discussed
in the main text
and detailed in the previous subsection. This effect is captured within the 
two-site calculation but is not considered in the single-site one.

The effect of bonding-antibonding splitting in the bandstructure can be 
detected via optical conductivity. Without considering the AFM exchange 
interaction between quasi-local moments, the strong band renormalization 
shifts the interband peak of the optical conductivity from about 
$1.3$-$1.5$ eV to about $0.7$ eV, as shown in the single-site results in 
Fig.~\ref{s7}{\bf b}.
By contrast, in the two-site calculation,
where the effects of AFM exchange interaction is considered, the interband peak 
is located at about $1$ eV
as shown in Fig.~4{\bf b} of the main text,
which is consistent with the experimental results.

\subsection{2.4 Similarity of the electron correlation effects between the 
	low-pressure 
	high-pressure phases}

\begin{figure}
	\includegraphics[width=0.7\textwidth]{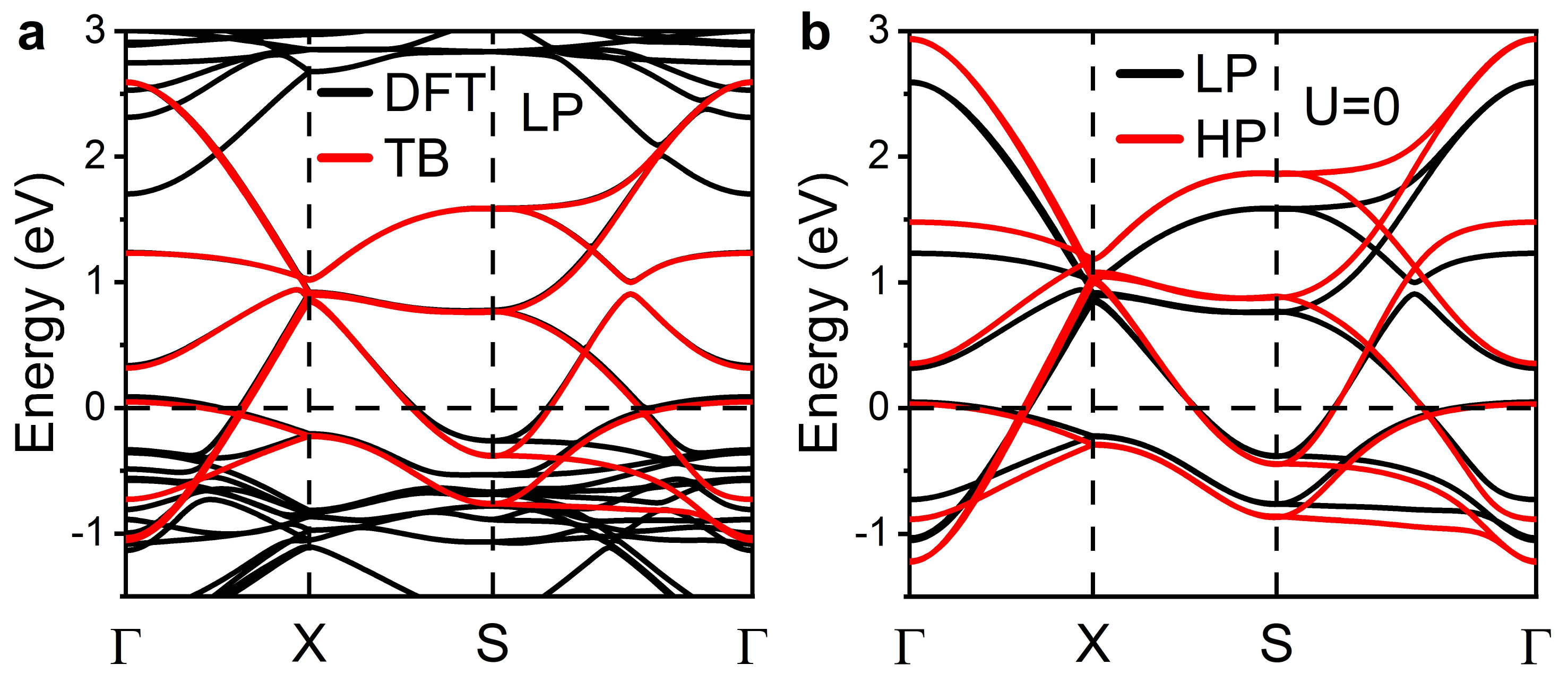}
	\caption{{\bf a} Comparison between the DFT bandstructure 
		at ambient pressure and the one calculated from the tight-binding model 
		for 
		the low-pressure phase. {\bf b} Comparison of the electronic bands of 
		the 
		low-pressure phase with the folded ones of the high-pressure phase at 
		$U=0$.}
	\label{s8}
\end{figure}

Our results in the main text are obtained based on a tight-binding model 
that describes the high-pressure phase. We expect our results to also describe 
the major features of the low-pressure phase.
To verify this, 
we performed the DFT calculation at ambient pressure and correspondingly 
constructed a tight-binding model for the low-pressure phase. As shown in 
Fig.~\ref{s8}{\bf a}, the band structure of this tight-binding model reproduces 
a similar one of DFT in the energy window of interest, from -1 to 
2.5 eV. We then compare the tight-binding models for the low-pressure and 
high-pressure phases. The major differences are as follows:
\begin{enumerate}
	\item The Ni-O-Ni angle between layers changes from $168^\circ$ in the 
	low-pressure phase to $180^\circ$ in the high-pressure phase; this 
	increases 
	the inter-layer hopping and the bonding-antibonding splitting.
	\item The applied pressure shortens the in-plane Ni-Ni length; this 
	increases 
	the in-plane hopping between the $x^2-y^2$ orbitals and increases the 
	bandwidths 
	of 
	the $x^2-y^2$ bands.
\end{enumerate}

\begin{figure}
	\includegraphics[width=0.7\textwidth]{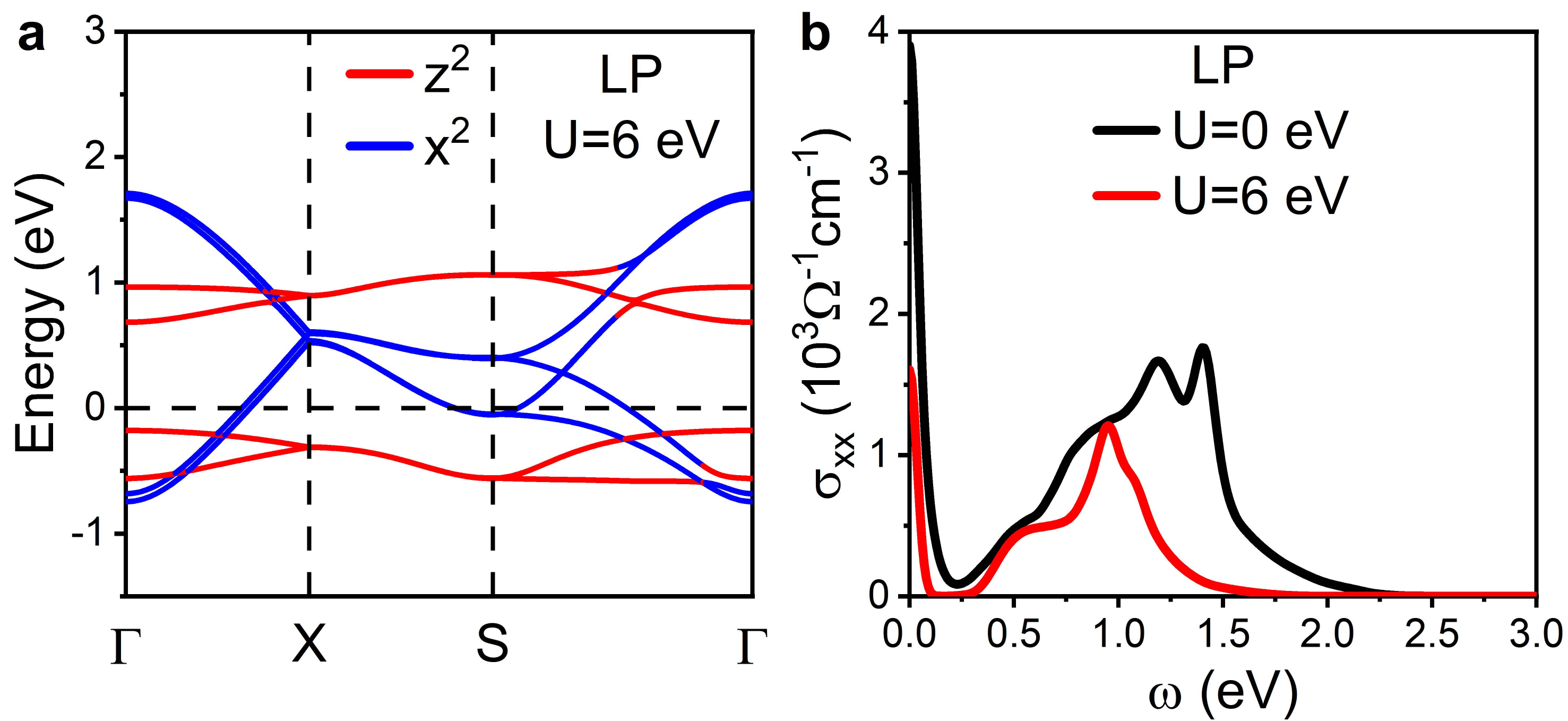}
	\caption{{\bf a} Electronic bandstructure of the model for 
		the low-pressure phase at $U=6$ eV. {\bf b} In-plane optical 
		conductivity of the model for  
		low-pressure phase 
		at $U=0$ and $U=6$ eV, respectively.}
	\label{s9}
\end{figure}

Fig.~\ref{s8}{\bf a} compares the electronic bands of the low-pressure 
phase 
and the folded bands of the high-pressure phase for $U=0$. Though minor 
differences do exist, these two models generate very similar bandstructures 
(after appropriate band folding).
We then study the correlated electron structure of the low-pressure model at  
finite $U$. As shown in Fig.~\ref{s9}{\bf a} and compared with Fig.~3{\bf b}, 
the low-pressure model exhibits similarly strong orbital-selective electron 
correlation effects as the high-pressure one, as expected. 
More importantly, we have examined that all the 
characteristics of the strong orbital-selective correlations, including the 
strong 
orbital-dependent band renormalization, the $\gamma$ band's sinking below-$E_F$,
the reduction of the
Drude weight, and the 1 eV interband
component in the optical conductivity 
(Fig.~\ref{s9}{\bf b}), are similar with the results of the high-pressure model 
discussed in the main text.

\end{document}